\def\b{\bibitem}

\def\be{\begin{equation}}
\def\ee{\end{equation}}
\def\bea{\begin{eqnarray}}
\def\eea{\end{eqnarray}}
\def\bml{\begin{mathletters}}
\def\eml{\end{mathletters}}

\documentstyle[aps,prl,eqsecnum,psfig,floats]{revtex}
\begin{document}
% Macros for the various macro package names, etc.
\def\SNG{{\em Physical Review Style and Notation Guide}}
\def\LUG {{\em \LaTeX{} User's Guide \& Reference Manual}}
\def\btt#1{{\tt$\backslash$\string#1}}%
\def\REVTeX{REV\TeX}
\def\AmS{{\protect\the\textfont2
        A\kern-.1667em\lower.5ex\hbox{M}\kern-.125emS}}
\def\AmSLaTeX{\AmS-\LaTeX}
\def\BibTeX{\rm B{\sc ib}\TeX}
%\makeatletter
%\tighten
\twocolumn[\hsize\textwidth\columnwidth\hsize\csname@twocolumnfalse%
\endcsname
 
\title{Reply to Comment on ``Absence of electron dephasing at zero 
       temperature''
       %\\by D.S. Golubev, A.D. Zaikin, and G. Sch{\"o}n, cond-mat/0111527
}
\author{T.R.Kirkpatrick}
\address{Institute for Physical Science and Technology, and Department of 
         Physics\\
         University of Maryland,\\ 
         College Park, MD 20742}
\author{D.Belitz}
\address{Department of Physics and Materials Science Institute\\
         University of Oregon,\\
         Eugene, OR 97403}
\date{\today}
\maketitle

\begin{abstract}
We explain why the objections raised by Golubev et al. 
in a comment on cond-mat/0111398 are not valid.

\end{abstract}
\pacs{PACS numbers:  }
]
%\narrowtext
%\tableofcontents

In a recent comment\cite{comment} on our paper\cite{us}, Golubev et al.
raise two objections to our proof that the Cooperon is massless
at zero temperature. They claim that, (1) the electron-electron interaction
breaks time-reversal invariance, and (2) our proof breaks down for the
case of a Coulomb interaction, as opposed to a short-ranged interaction.
We disagree with both of these arguments.

Concerning (1), it is well established that interacting many-particle
systems, classical or quantum, are time reversal invariant in equilibrium
in the absence
of external magnetic fields or magnetic impurities\cite{KadanoffMartin}. 
Time reversal is an
exact symmetry of our starting Hamiltonian, as well as the one studied
by the authors of Ref.\ \onlinecite{comment}. This leads to the exact
symmetry of the Green function used in Ref.\ \onlinecite{us}, which in
turn relates the diffuson and Cooperon, respectively. To avoid 
misunderstandings we stress again, as we have in Ref.\ \onlinecite{us},
that this relation holds for the disconnected (with respect to the
interaction) parts of the particle-hole (``diffuson'') and 
particle-particle (``Cooperon'') correlation functions 
only\cite{Raimondi_etal}.

Concerning (2), there is no difference in the behavior of systems
interacting via a Coulomb interaction and a short-ranged interaction,
respectively, in dimensions $d>2$. Furthermore, it has been shown in
Ref.\ \onlinecite{us_fermionsII} how to generalize our Ward identity
to deal directly with a bare Coulomb interaction, if this is desired.
This procedure is substantially more involved than the
simple limit taken in Ref.\ \onlinecite{comment}.

In summary, the objections raised by Golubev et al. are not valid.

\acknowledgments
This work was supported by the NSF under Grant Nos. DMR-99-75259 and
DMR-98-70597.

\end{document}